# From ripples to spikes: a hydro-dynamical physical mechanism to interpret femtosecond laser induced self-assembled structures

George D. Tsibidis[1,♣], C. Fotakis[1,2], E. Stratakis[1,♣]

[1]*Institute of Electronic Structure and Laser (IESL), Foundation for Research and Technology (FORTH), N. Plastira 100, Vassilika Vouton, 70013, Heraklion, Crete, Greece*

[2]*Physics Department, University of Crete, Heraklion 71409, Crete, Greece*

Materials irradiated with multiple femtosecond laser pulses in sub-ablation conditions are observed to develop various types of self-assembled morphologies that range from nano-ripples to periodic micro-grooves and quasi-periodic micro-spikes. Here, we present a physical scenario that couples electrodynamics, describing surface plasmon excitation, with hydrodynamics, describing Marangoni convection and counter-rolls, to elucidate this important sub-ablation regime of light matter interaction in which matter is being modified, however, the underlying process is not yet fully understood. The proposed physical mechanism could be generally applicable to practically any conductive material structured by ultrashort laser pulses, therefore it can be useful for the interpretation of further critical aspects of light matter interaction.

[♣] E-mail: tsibidis@iesl.forth.gr ; stratak@iesl.forth.gr

Over the past decades, material processing with femtosecond (fs) pulsed lasers has received considerable attention due to its multiple diverse applications ranging from micro-device fabrication to optoelectronics, microfluidics and biomedicine [1-4]. Direct fs laser surface micro- and nano- patterning has been demonstrated in many types of materials including semiconductors, metals, dielectrics, ceramics, and polymers. Following irradiation with ultrashort pulses, multiphysical phenomena take place [1, 5, 6] while a plethora of surface structures can be realized. In case of metallic and semiconducting materials, including Silicon (Si) shown in Fig. 1, periodic submicron ripples are formed at low (Fig.1a,b), microgrooves at intermediate (Fig.1c,d) [7] and quasi-periodic arrays of microspikes (Fig.1e,f) [8, 9] at high number of pulses (*NP*). Similar-looking groove- and spiky-like macro-structures can also be found in nature, when large thermal gradients are developed, for example, in the formation of penitentes [10], jet-like feature in the oceans [11], volcano lava laminar flow [12].

To date, there are numerous reports on the mechanism behind morphological changes [13-24] and, more specifically, formation of sub-wavelength sized ripples [25-31]. While a thermocapillary mechanism of melt flow due a temperature gradient has been suggested in previous works [24, 32] to describe morphological changes for nanosecond pulses, there is no interpretation of the microgrooves generation (that develop perpendicularly to the ripples and have a substantially larger periodicity than the ripples) and transition to spike formation from the viewpoint of physics; furthermore, while the excitation of surface plasmon (SP) wave mechanism has been considered as the most prominent scenario [25-31] for the ripple formation for fs pulses, no previous investigation has addressed the possibility of suppression of electrodynamical mechanisms (i.e. SP excitation) with increasing irradiation and how it could influence the surface profile of the material; hence a unified theoretical model that could provide a detailed description of the underlying fundamental physical processes for the entire range of structures is still elusive. To explain the laser-induced formation of such microstructures, a variety of complex mechanisms should be explored, including electrodynamics, phase transitions and molten material hydrodynamics as well as investigation of the role of viscous materials dynamics in determining the shape of the surface profile.

In this Letter, we present a theoretical model to account for the whole range of structures formed upon the interaction of a fs laser beam with a conductive surface, shown for Si in Fig 1. Specifically the critical role of hydrodynamics on the fs surface structuring process is elucidated. The developed theoretical model comprises [33] (i) an electromagnetics component that describes the surface plasmon wave (SPW) excitation and interference with the incident beam that leads to a periodic modulation of the laser field energy density, (ii) a heat transfer component that accounts for carrier-lattice thermalisation through particle dynamics and heat conduction and carrier-phonon coupling, and (iii) a hydrodynamics component that describes the molten material dynamics and re-solidification process assuming an incompressible Newtonian fluid flow that includes recoil pressure and surface tension contributions [30] as well as Marangoni effects [34, 35] and hydrothermal convection [36].

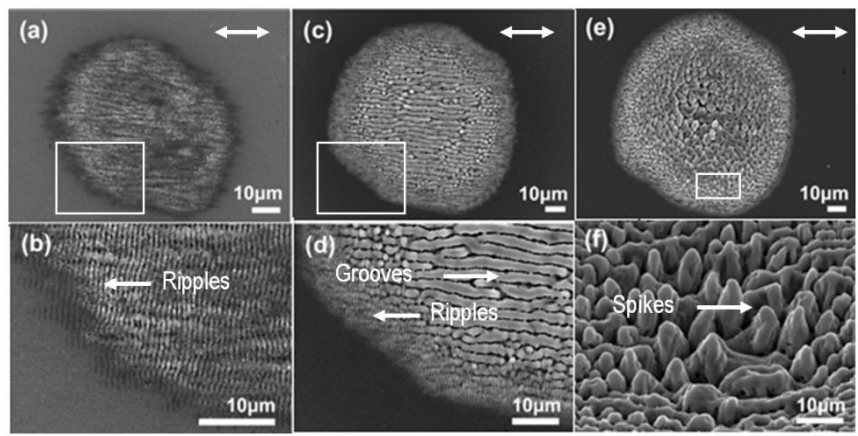

FIG.1. Morphological changes induced on a *Si* surface following irradiation with $NP=10$ (a), 40 (c), 100 (e) pulses, where (b), (d), (f) provide an enlarged area ($E_d=0.7J/cm^2$, $\tau_p=430fs$, $R_0=15\mu m$. Double-ended arrows indicate the laser beam polarisation).

To predict the laser-induced morphological changes, simulations were performed with a *p*-polarised, Gaussian (both spatially and temporally) laser pulse of fluence $E_d=0.7J/cm^2$, pulse duration $\tau_p=430fs$, irradiation spot radius $R_0=15\mu m$ and laser beam wavelength $\lambda=800nm$ [30]. A systematic analysis of the fundamental mechanism reveals that mass removal due to the fact that excessive temperatures are reached (larger than ~$0.90T_c$, $T_c=3540K$) [37] and inhomogeneous deposition of the laser energy leads to recoil pressure, surface tension variance and temperature gradients (Fig.1a in[33]). These effects induce a Marangoni-driven flow and capillary waves that eventually leads upon resolidification to the formation of a crater and a rippled profile while mass conservation and surface tension forces produce a protrusion near the periphery of the spot (Fig.2a,b in [30, 33]). To evaluate the dependence of incubation effects with increasing number of pulses, the inhomogeneous energy deposition into the irradiated material is computed through the calculation of the product $\eta(\mathbf{k},\mathbf{k}_i) \times |b(\mathbf{k})|$ as described in the Sipe-Drude model [38, 39]. In the above expression, $\eta$ describes the efficacy with which the surface roughness at the wave vector $\mathbf{k}$ (i.e. $|\mathbf{k}|=2\pi/\lambda$) induces inhomogeneous radiation absorption, $\mathbf{k}_i$ is the component of the wave vector of the incident beam on the material's surface plane and $b$ represents a measure of the amplitude of the surface roughness at $\mathbf{k}$. The efficacy factor depends on the dielectric constant $\varepsilon$ of the material [39] and therefore any variation of the carrier temperature and density due to inhomogeneous radiation absorption is expected to influence the dielectric constant, SPW and finally ripple periodicity. Simulation results and experimental data [30] indicate the production of subwavelength size ripples (Fig.1a in [33]) which is consistent with previous experimental observations [7] and theoretical predictions [39].

To investigate morphological changes for a larger number of pulses, we notice that upon irradiation of the modified material with $NP \geq 20$, a new type of laser-induced pattern is observed whose formation process has not been understood to date (Fig.1c,d). Interestingly, the produced structures exhibit a spatial periodicity which is higher than the ripples' wavelength (i.e. more than twice the laser wavelength). To explore the underlying physical mechanism that leads to the grooves formation, we firstly quantify the inhomogeneous energy deposition through the computation of the efficacy factor. This is presented in Figure 2a, showing the dependence of the ripples periodicity and the maximum efficacy factor on the carrier density. As can be seen, for carrier densities, ranged from $N_{max}= 4.83 \times 10^{21}$ cm$^{-3}$ (i.e. where efficacy factor attains maximum value) to $N_{cr}=4.29 \times 10^{21}$ cm$^{-3}$ corresponding to the minimum value for SPW excitation (i.e. $Re(\varepsilon)=-1$), the efficacy factor is decreasing that also leads to a decreasing inhomogeneous energy deposition [38, 39]. On the other hand, for carrier densities higher than $N_{max}$ an abrupt decrease and eventual vanishing of the efficacy factor occurs [25]. Furthermore, upon increasingly irradiating the material with more pulses, the surface relief deepens, characterized by an increasingly smaller shape factor $s$, which also leads to a rapidly decreasing efficacy factor [33]. The latter is nevertheless expected as energy absorption is influenced by the variations in the local angle of incidence and reflectivity. As a result, a gradual suppression of ripples would follow as the diminishing efficacy factor on the one hand leads to an amplitude decrease of the produced ripples while the efficacy factor peak is reached at a level which is lower than the threshold of SPW excitation.

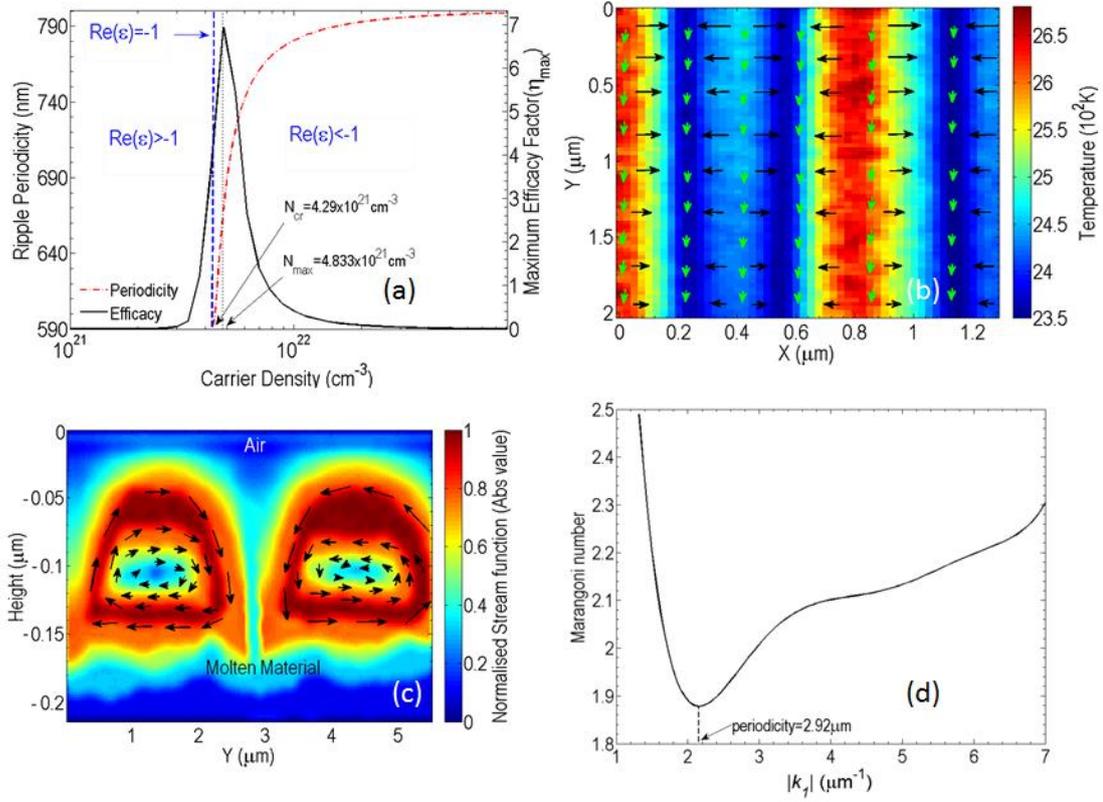

FIG.2 (Color Online). (a) Ripple periodicity dependence on the number of excited carriers (red *dotted* line). Maximum efficacy factor variation (for *s*=0.4) with increasing density number is indicated by *solid* line. Vertical blue and brown *dashed* lines indicates the condition for SPW excitation ($Re(\varepsilon)<-1$) and maximum $N_{max}$, respectively, (b) Temperature field on the surface at *t*=1ns (*black* and *green* arrows indicate the flow movement along X-, Y- directions), (c) Stream function field at *t*=1ns for *NP*=20 along Y-axis (*Black* arrows indicate direction of the fluid movement considering the *absolute* value of the stream function [33] normalised to 1), (d) Marginal stability of the hydrothermal patterns for *NP*=20 occurs at $|k_1|$=2.1528μm$^{-1}$ (period=2.92μm) ($E_d$=0.7J/cm$^2$, $\tau_p$=430fs, $R_0$=15μm).

Ripples' suppression suggests that the origin of grooves cannot be ascribed to electrodynamics and SPW excitation; by contrast, a thorough investigation of the transient lattice temperature profile and the subsequent dynamic flow of the resulting molten material indicate that hydrodynamical effects are behind the grooves formation. We first elaborate on the dynamics of the molten material and analyse the melt resolidification process. More specifically, the rippled profile created for low number of pulses constitutes a grating structure with a substantially larger surface modulation across the ripples (*black* arrows in Fig. 2b). However, lateral fluid movement from both sides of the well (*blue* stripes in Fig. 2b) will hinder further material displacement. By contrast, the presence of a temperature gradient along the *Y*-axis leads to surface stress that drives a shear flow while the resulting temperature profile is capable of destabilizing the molten layer to Marangoni convection and production of hydrothermal waves perpendicularly to the thermal gradient [36]. This thermal-convective instability is indicative for fluids characterised by a low Prandtl number *Pr* (i.e. *Pr* (for molten Si) =0.017 [33]) leading to transverse counter-rotating rolls (Fig.2c) [33, 36, 40] in contrast to longitudinal convection rolls that occur for liquids with high Prandtl number [36, 41]. We hence postulate that when the temperature difference within the melt exceeds a critical value, convection rolls generated by surface tension gradients and hydrothermal waves will induce surface tractions, developed perpendicular to ripples orientation. In order to verify the above phenomenological description of the melt hydrodynamic flow and predict the generation of hydrothermal waves in thin liquid films through heat transfer and Marangoni convection, we solve Eqs.(7-12, 17-24) (in [33]) and estimate the conditions that will lead to the formation of the aforementioned transverse structures [36, 40-42]. In principle, the Marangoni number, *M*, that represents the ratio of rate of convection and rate of conduction, is used to characterize the flow due to surface tension gradients through the relation $M=(\partial\sigma/\partial T)\Delta T\Delta H/(\alpha\mu)$, where $\partial\sigma/\partial T$, $\Delta T$, $\Delta H$, $\alpha$, and $\mu$ are the surface tension gradient, average temperature difference between the lower and upper layer of the liquid, average thickness of the fluid, thermal diffusivity and dynamic viscosity of the molten material. However, the fluid flow in a curved/inclined space and the spatio-temporal dependence of the rest of the parameters due to the highly nonlinear nature of the problem does not allow the derivation of a simple, analytic expression that provides a value for the onset of the roll movement, their stability conditions and final profile upon solidification. Nevertheless, a numerical solution of the aforementioned equations is pursued using a finite difference methodology and a staggered grid [30, 33] in which stress-free and no-slip (i.e. $\boldsymbol{u}=0$) boundary conditions are imposed on the liquid-solid interface while shear stress boundary conditions are assumed on the free surface. Using a common approach, the first step is to find equations describing the overall convective flow (i.e. base flow) followed by a linear stability analysis of small perturbations to the flow [43]. Then, the revised equations are solved using a semi-implicit time discretization fixing a wave vector $\boldsymbol{k_n}=\left(k_x(n),0,k_z(n)\right)$ and seeking solutions in the form of normal modes for $u_{x,z}(\vec{r},t) \sim W(z)e^{-\omega t}\sum_n c_n e^{i\vec{k}_n\bullet\vec{r}}$,

$T(\vec{r},t) \sim \theta(z)e^{-\omega t}\sum_n c_n e^{i\vec{k}_n \cdot \vec{r}}$ and $P(\vec{r},t) \sim p(z)e^{-\omega t}\sum_n c_n e^{i\vec{k}_n \cdot \vec{r}}$, for the velocity, temperature and pressure, respectively, where $n=1$ corresponds to the formation of transverse roll structures and $\omega$ ($<0$) is the growth rate of the structures. Based on the values of the hydrodynamic parameters and the temperature dependence of the physical parameters of liquid Si for $NP=20$, counter rotating vortex rolls are formed which propagate perpendicularly to the molten material movement (Fig.2c). The molten flow is characterised by decreasing vorticity inside the roll while the stream function at the bottom of the liquid indicates an almost parallel flow (that diminishes at larger depths) with shear (see also Fig.4b in [33]). Furthermore, the development of two rolls with opposite flow direction leads to a rapidly decreasing (i.e. close to stagnation) speed at the meeting point. To determine the stability of the roll patterns, we compute the (spatial) average Marangoni number as a function of $|k_1|$ to estimate the critical value of the onset of convection which yields a periodicity equal to $|k_1|/(2\pi)$ ~2.92μm (Fig.2d). The graph suggests that the critical value of the Marangoni number $M_c$~1.88 is the minimal value for which the molten layer is convectively unstable and leads to the formation of rolls with direction parallel to the ripples [43].

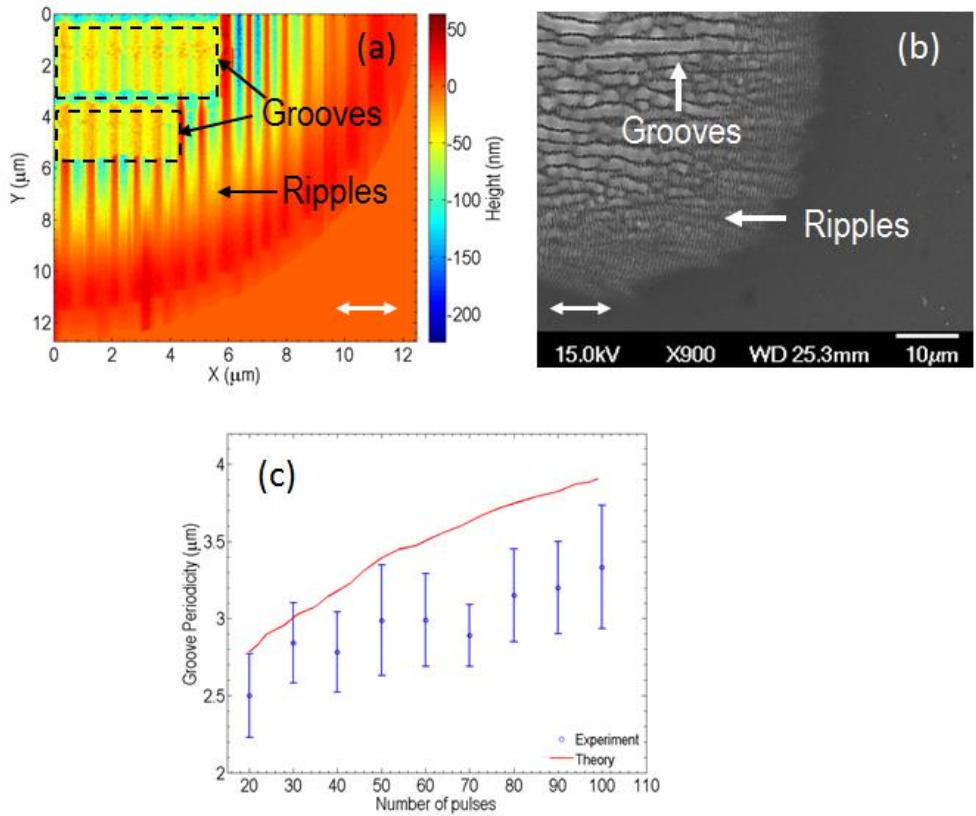

FIG.3 (Color Online). (a) Groove-ripple pattern for $NP$=20, (b) Experimental results for $NP$=20, (c) Groove periodicity vs. number of pulses. ($E_d$=0.7J/cm$^2$, $\tau_p$=430fs, $R_0$=15μm). (Double-ended arrow indicates the laser beam polarisation).

To predict the surface modification of the material following the fluid movement described above, we investigate the morphology attained when resolidification takes place for conditions that lead to $M_c$~1.88. The resulting surface profile is illustrated in Fig.3 (upper view) for $NP$=20 with a groove periodicity equal to $\Lambda_{groove}$=2.79μm (i.e. resolidification process leads to the reduction of the groove periodicity compared to the roll structure wavelength of 2.92 μm); it is obvious that three types of structures are produced at the same time: (i) conventional ripples with a subwavelength periodicity in the spot area where the criteria for SPW excitation are fulfilled, (ii) grooves (area enclosed in *dashed* lines) with periodicity of ~$\Lambda_{groove}$, and (iii) pseudoripples (to distinguish them from the conventional ripples) appearing on top of the grooves as well as in the spaces among them and they are remains of the ripples formed at low $NP$s (see [33]). Therefore the amplitude of the capillary waves created due to Marangoni convection is not large enough to eliminate the ripple profile leading to the formation of pseudoripples. An experimental validation of the grooves and ripples coexistence and their relative unequal periodicities is illustrated in Fig.3b. While the pseudoripples do not appear in this case, unlike theoretical results (Fig.3a), experiments performed on metal surfaces (i.e. Titanium [33]) show that the presence of such structures is possible and the presented scenario could reveal the critical role of hydrodynamic instabilities on surface patterning. Numerical simulations indicate that the height of the pseudoripples decreases with increasing $NP$ due to the increase of the volume of the elevated molten material leading to progressive suppression of capillary wave amplitude [33]. Further analysis was also conducted to investigate grooves' periodicity dependence on the number of pulses (100≥$NP$≥20). Upon increasing $NP$, the enhanced mass removal and deepening of the surface profile causes enhanced energy deposition and temperature gradients. As a result, an increased outward movement of the molten is produced. Accordingly, simulation results predict that as $NP$ increases hydrothermal rolls with an increasing periodicity patterns are produced. Hence, the proposed model predicts that the groove periodicity is an increasing function of the number of pulses which is similar to the trend observed in the experiments (Fig.3c).

To explore whether the gradual diminishing of the pseudoripples and increase of grooves periodicity further influences the melt flow and the resulting profile, we performed simulations at higher number of pulses, i.e. $NP \geq 100$. The disappearance of pseudoripples combined with the enhanced temperature gradients generated from the grooves' grating in this case drives predominantly the fluid flow in a direction perpendicularly to laser beam polarization vector (Fig.4a). Hence, surface tension gradients and temperature differences will produce an overall preferred Marangoni convection along the *X*-axis and thus hydrothermal waves will be induced, following a similar process to the one described previously for grooves formation (however, the rolls are currently developed in the opposite direction). In particular, base flow with a subsequent linear analysis procedure is followed for a wavevector $\boldsymbol{k_n} = (0, k_y(n), k_z(n))$, while the onset for the development of roll structures occurs on top of the original (i.e. along the *X*-axis) grooves when the wavelength of the periodical modulation is ~2.3μm corresponding to a Marangoni number equal to $M_c$~2.4 (Fig.4b). Notably, the roll patterns due to the convective instability evolve into protruding structures with a periodicity of $\Lambda_{protruded}$ =2.1μm upon melt solidification. The height of the protruding structures progressively increases with increasing the number of pulses giving rise to spike-like assemblies upon further irradiation (Fig.4c encircled objects) [33]. The simulation predictions presented above show remarkable similarity with the surface morphology obtained from experiments performed at identical irradiation conditions, while the produced hills can be viewed as the precursors of microspikes. This is evident in Fig. 4d, where the coexistence of grooves and microspikes can be observed. Moreover, it is obvious that the grooves' profile is spatially modulated and the corrugation wavelength is of the size of the grooves' periodicity (≥2.1μm). The similarities between our theoretical model and experimental findings (Fig.4c,d), especially concerning the predicted periodicities could work towards formulating a comprehensive theoretical framework that accounts for the formation of these structures. Although further investigation is required to establish a more enlightening picture of the process of microgrooves and spikes formation, it is evident that the proposed model underlines conclusively the predominant role of hydrodynamical processes.

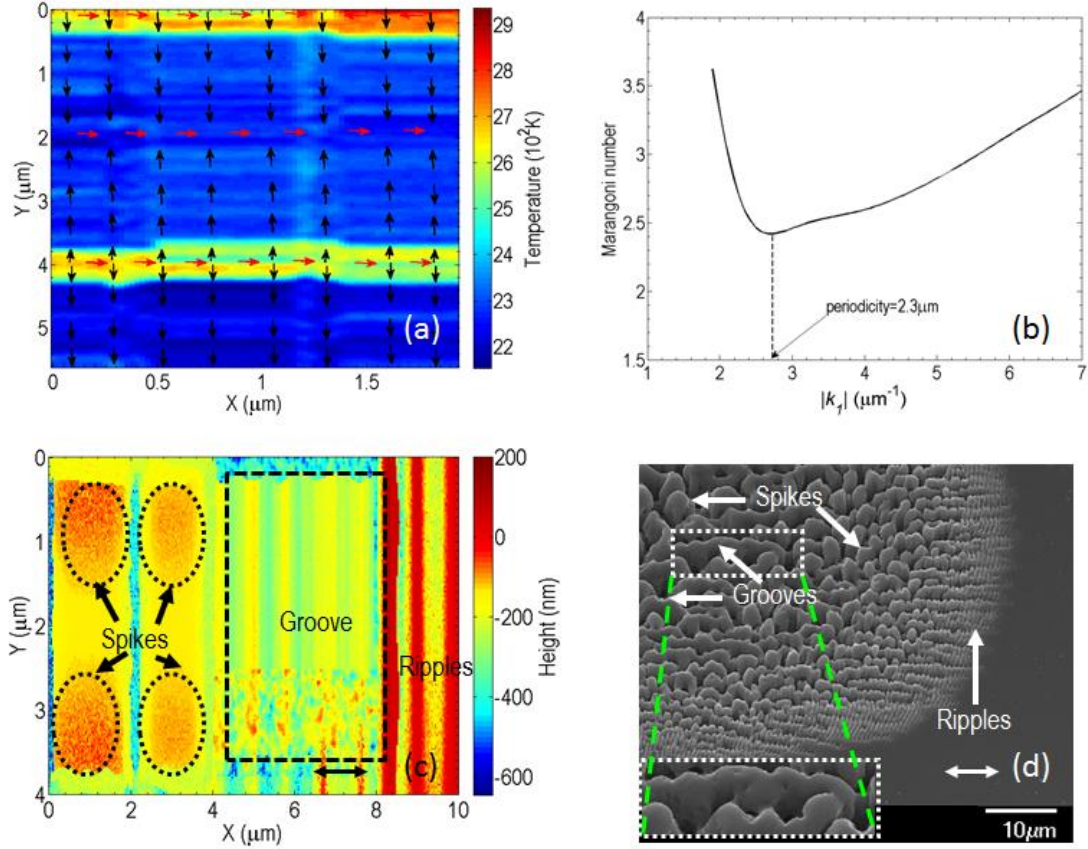

FIG.4 (Color Online). (a) Temperature field on the surface at $t$=1ns for $NP$=120 (*black* and *red* arrows indicate the flow movement along $X$-, $Y$- directions while the *blue* region characterises the area occupied by the groove), (b) Stability of the convection roll patterns for $NP$=120 occurs at $|k_1|$=2.7318μm$^{-1}$ (period=2.3μm), (c) Surface pattern for $NP$=120 (encircled objects are hills and groove is enclosed in the boxed area), (d) Experimental results for $NP$=200 ($E_d$=0.7J/cm$^2$, $\tau_p$=430fs, $R_0$=15μm) (Double-ended arrow indicates the laser beam polarisation).

In summary, the physical mechanism that accounts for the various stages of the structuring process taking place upon irradiation of a conductive solid with multiple identical fs laser pulses has been demonstrated. In particular, a hybrid model is proposed based on melt hydrodynamics complemented with electromagnetic interference effects to account for ripples, while Marangoni shear generated convection dominates the mechanism that leads to hydrothermal waves and eventually to microgrooves and microspikes formation. The proposed interpretation can firstly enhance our understanding of the fundamental physical processes that characterise ultrashort laser-matter interaction and secondly allow control of the optoelectronic characteristics of the different structures, which is expected to meet a wide range of applications in industry and biomedicine.

The authors acknowledge support from the *'3DNeuroscaffolds'* research project.

**Supplementary Material for 'From ripples to spikes: a hydro-dynamical physical mechanism to interpret femtosecond laser induced self-assembled structures'**

George D. Tsibidis[1,♣], C. Fotakis[1,2], and E. Stratakis[1,♣]

[1]*Institute of Electronic Structure and Laser (IESL), Foundation for Research and Technology (FORTH), N. Plastira 100, Vassilika Vouton, 70013, Heraklion, Crete, Greece*

[2]*Physics Department, University of Crete, Heraklion 71409, Crete, Greece*

---

[♣] E-mail: tsibidis@iesl.forth.gr; stratak@iesl.forth.gr

1. **Theoretical formulation**

Below we present the fundamental theoretical concept that comprises thermal and hydrodynamical components to describe the response of the material after irradiation with ultrashort pulses.

Ultrashort-pulsed lasers first excite the charge carriers (electron-hole pairs) in semiconductors while their energy is subsequently transferred to the lattice. The relaxation time approximation to Boltzmann's transport equation [44] is employed to determine the carrier density number, carrier energy and lattice energy. The model assumes an equal number of electron and holes in the solid and no electron photoemission is considered after the laser irradiation [45]. The evolution of the carrier density number $N$, carrier temperature $T_c$ and lattice temperature $T_l$ are derived using the carrier, carrier energy and lattice heat balance equations. Based on this picture the following set of equations determine the temperature and particle dynamics [44, 46-48]

$$C_c \frac{\partial T_c}{\partial t} = \vec{\nabla} \bullet \left((k_e + k_h)\vec{\nabla} T_c\right) - \frac{C_c}{\tau_e}(T_c - T_l) + S(\vec{r},t)$$

$$C_l \frac{\partial T_l}{\partial t} = \vec{\nabla} \bullet \left(K_l \vec{\nabla} T_l\right) + \frac{C_c}{\tau_e}(T_c - T_l)$$

$$\frac{\partial N}{\partial t} = \frac{\alpha}{h\nu}\Omega I(\vec{r},t) + \frac{\beta}{2h\nu}\Omega^2 I^2(\vec{r},t) - \gamma N^3 + \theta N - \vec{\nabla} \bullet \vec{J} \quad (1)$$

$$\Omega = \frac{1 - R(T_l)}{\cos\varphi}$$

The laser intensity in Eqs.(1-2) is obtained by considering the propagation loss due to one-, two- photon and free carrier absorption [44]

$$\frac{\partial I(\vec{r},t)}{\partial z} = -(\alpha + \Theta N)I(\vec{r},t) - \beta I^2(\vec{r},t) \quad (2)$$

assuming that the laser beam is Gaussian both temporally and spatially and the transmitted laser intensity at the incident surface is expressed in the following form [48]

$$I(r,z=0,t) = \frac{2\sqrt{\ln 2}}{\sqrt{\pi}\tau_p} E_p e^{-\left(\frac{2r^2}{R_0^2}\right)} e^{-4\ln 2\left(\frac{t-t_0}{\tau_p}\right)^2} \quad (3)$$

where $E_p$ is the fluence of the laser beam and $\tau_p$ is the pulse duration (i.e. full width at half maximum), $R_0$ is the irradiation spot-radius (distance from the centre at which the intensity drops to $1/e^2$ of the maximum intensity.

In the present work, we assume conditions that lead to temperatures greater than ~$0.90T_{cr}$ (for silicon, $T_{cr}=5159^0K$ [49]) for a small part of the material. While this portion is evaporated, a superheated liquid still remains in the system which undergoes a slow cooling. Furthermore, to introduce the phase transition that causes the bulk temperature to exceed the silicon melting temperature $T_m$ (~1687 $^0K$), the second equation in Eq.1 has to be modified properly to include the phase change in the sold-liquid interface

$$\left(C_l \pm L_m \delta(T_l - T_m)\right)\frac{\partial T_l}{\partial t} = \vec{\nabla} \cdot \left(K_l \vec{\nabla} T_l\right) + \frac{C_c}{\tau_e}(T_c - T_l) \quad (4)$$

where $L_m$ is the latent heat of fusion. A suitable representation of the $\delta$-function should accomplish (for numerical calculations) a smooth transition between solid and liquid phases [50], therefore the following expression is used

$$\delta(T_l - T_m) = \frac{1}{\sqrt{2\pi}\Delta} e^{-\left[\frac{(T_l-T_m)^2}{2\Delta^2}\right]} \quad (5)$$

where $\Delta$ is in the range of 10-100$^0K$ depending on the temperature gradient. The sign in front of the term that contains the $\delta$-function depends on whether melting or solidification takes place. One aspect that should not be overlooked is that melting of silicon not only induces a solid-to-liquid phase transition, but also alters its properties since the molten material exhibits metal behaviour. Hence, a revised two-temperature model that describes heat transfer from electrons to lattice has to be employed [51] and thereby, for temperatures above $T_m$, Eqs.1 need to be replaced by the following two equations that describe electron-lattice heat transfer

$$C_e \frac{\partial T_e}{\partial t} = \vec{\nabla} \cdot (K_e \vec{\nabla} T_e) - \frac{C_e}{\tau_E}(T_c - T_L)$$

$$C_L \frac{\partial T_L}{\partial t} = \frac{C_e}{\tau_E}(T_c - T_L) \tag{6}$$

where $C_e$ and $C_L$ are the heat capacity of electrons and lattice (liquid phase), $K_e$ is the thermal conductivity of the electrons, while $\tau_E$ is the energy relaxation time for the liquid phase. The governing equations for the incompressible Newtonian fluid flow and heat transfer in the molten material are defined by the following equations:

(i). for the mass conservation (incompressible fluid):

$$\vec{\nabla} \cdot \vec{u} = 0 \tag{7}$$

(ii). for the energy conservation

$$C_L \left( \frac{\partial T_L}{\partial t} + \vec{\nabla} \cdot (\vec{u} T_L) \right) = \vec{\nabla} \cdot (K_L \vec{\nabla} T_L) \tag{8}$$

where $K_L$ is the thermal conductivity of the lattice. The presence of a liquid phase and liquid movement requires a modification of the second of Eq.8 to incorporate heat convection. Furthermore, an additional term is presented in the equation to describe a smooth transition from the liquid-to-solid phase (i.e. it will help in the investigation of the resolidification process)

$$C_L \left[ \frac{\partial T_L}{\partial t} + \vec{\nabla} \cdot (\vec{u} T_L) \right] - L_m \delta(T_L - T_m) \frac{\partial T_L}{\partial t} = \vec{\nabla} \cdot (K_L \vec{\nabla} T_L) \tag{9}$$

(iii). for the momentum conservation:

$$\rho_L \left( \frac{\partial \vec{u}}{\partial t} + \vec{u} \cdot \vec{\nabla} \vec{u} \right) = \vec{\nabla} \cdot \left( -P\mathbf{1} + \mu(\vec{\nabla} \vec{u}) + \mu(\vec{\nabla} \vec{u})^T \right) \tag{10}$$

where $\vec{u}$ is the velocity of the fluid, $\mu$ is the liquid viscosity, $P$ pressure. $C_L$ and $K_L$ stand for the heat capacity and thermal conductivity of the liquid phase, respectively. It is evident that the transition between a purely solid to a completely liquid phase requires the presence of an intermediate zone that contains material in both phases. In that case, Eq.10 should be modified accordingly to account for a liquid-solid two phase region (i.e. mushy zone) where the total velocity in a position should be expressed as a combination of the fraction of the mixtures in the two phases [52]. Nevertheless, to avoid complexity of the solution of the problem and given the small width of the two phase region with respect to the size of the affected zone a different approach will be pursued where a mushy zone is neglected and transition from to solid-to-liquid is indicated by a smoothened step function of the thermophysical quantities. Furthermore, as it will be explained in the Simulation section, the flow will be assumed to occur for two liquids of significantly different viscosity.

Vapour ejected creates a back (recoil) pressure on the liquid free surface which in turn pushes the melt away in the radial direction. The recoil pressure and the surface temperature are usually related according to the equation [53, 54]

$$P_r = 0.54 P_0 \exp\left( L_v \frac{T_L^S - T_b}{R T_L^S T_b} \right) \tag{11}$$

where $P_0$ is the atmospheric pressure (i.e. equal to $10^5$ Pa), $L_v$ is the latent heat of evaporation of the liquid, $R$ is the universal gas constant, and $T_L^s$ corresponds to the surface temperature. Given the radial dependence of the laser beam, temperature decreases as the distance from the centre of the beam increases; at the same time, the surface tension in pure molten silicon decreases with growing melt temperature (i.e $d\sigma/dT<0$), which causes an additional depression of the surface of the liquid closer to the maximum value of the beam while it rises elsewhere. Hence, spatial surface tension variation induces stresses on the free surface and therefore a capillary fluid convection is produced. Moreover, a precise estimate of the molten material behaviour requires a contribution from the surface tension related pressure, $P_\sigma$, which is influenced by the surface curvature and is expressed as $P_\sigma = K\sigma$, where $K$ is the free surface curvature. The role of the pressure related to surface tension is to drive the displaced molten material towards the centre of the melt and restore the morphology to the original flat surface. Thus, pressure equilibrium on the material surface implies that the pressure in Eq.10 should outweigh the accumulative effect of $P_r + P_\sigma$.

As the material undergoes a solid-to-liquid-to-solid phase transition, it is important to explore the dynamics of the distribution of the depth of the molten material and the subsequent surface profile change when solidification terminates. The generated ripple height is calculated from the Saint-Venant's shallow water equation [55]

$$\frac{\partial H(\vec{r},t)}{\partial t} + \vec{\nabla} \bullet \left( H(\vec{r},t)\vec{u} \right) = 0 \tag{12}$$

where $H(\vec{r},t)$ stands for the melt thickness. Hence, a spatio-temporal distribution of the melt thickness is attainable through the simultaneous solution of Eqs.(1-12).

Due to an inhomogeneous deposition of the laser energy on the semiconductor as a result of the exposure to a Gaussian-shape beam, the surface of material is not expected to be perfectly smooth after resolidification; further irradiation of the non-planar profile will give rise to a surface scattered wave [56]. According to theoretical predictions and experimental studies, the interference of the incident and the surface wave results in the development of periodic 'ripples' with orientation perpendicular ($p$-polarisation) to the electric field of the laser beam [56-58]. A revised process which that leads to the formation of the surface wave has been also proposed that involves surface plasmons [7, 26] where the ripple periodicity is provided by the expression $\lambda/(\lambda/\lambda_s \pm sin\varphi)$. The involvement of surface plasmon wave related mechanism in the generation of ripples will be employed in this work as the metallic behaviour of silicon at large temperatures allows excitation of surface plasmon waves. The plasmon wavelength, $\lambda_s$, is related to the wavelength of the incident beam through the relations [26]

$$\lambda_s = \lambda / \eta,$$
$$\eta = \text{Re}\left(\frac{\varepsilon' + \varepsilon_d}{\varepsilon' \varepsilon_d}\right)^{1/2} \tag{13}$$
$$\varepsilon' = \left(1 + (\varepsilon_g - 1)\left(1 - \frac{N}{n_0}\right) - \frac{N}{N_{cr}} \frac{1}{\left(1 + i\frac{1}{\omega\tau_e}\right)}\right)$$

where $\varepsilon_d$ ($\varepsilon_d =1$) is the dielectric constant of air, $\varepsilon_g$ stands for the dielectric constant of unexcited material ($\varepsilon_g=13.46+i0.048$), $\omega$ is the frequency of the incident beam, $n_o$ is the valence band carrier density ($n_o=5\times10^{22}$cm$^{-3}$), $N_{cr}=m_{eff}\varepsilon_0\omega^2/e^2$ where $m_{eff}$ is the effective electron mass [59].

Although the equations presented in the previous section are still valid, some modification has to be performed to the form of the laser intensity beam due to the interference of the incident and the surface plasmon wave. The final intensity on the surface is provided by the following expression

$$I_{surf}(\vec{r},t) = \left\langle \left|\vec{E}_i + n_{mat}\vec{E}_s\right|^2 \right\rangle e^{-\left(\frac{2r^2}{R_0^2}\right)} e^{-4\ln 2 \left(\frac{t-t_0}{\tau_p}\right)^2} \quad (14)$$

where $n_{mat}$ is the refractive index of silicon, $\vec{E}_i = \vec{E}_{i,o}(\vec{r})\exp(-i\omega_i t + i\vec{k}_i \cdot \vec{r})$ and $\vec{E}_s = \vec{E}_{s,o}(\vec{r})\exp(-i\omega_s t + i\vec{k}_s \cdot \vec{r})$ and $\omega_i$, $\omega_s$ are the frequencies of the incident beam (equal to $\omega$) and the surface plasmon wave, respectively. The magnitude of the electric field of the incident wave can be calculated by the expression

$$\frac{E_d}{\tau_p} = \frac{c\varepsilon_0\sqrt{\pi}\left\langle\left|\vec{E}_i\right|^2\right\rangle}{2\sqrt{\ln 2}} \quad (15)$$

while the magnitude of the electric field of the longitudinal surface plasmon wave [60] is taken to be of the order of the electric field of the incident wave.

2. **Simulation**

The proposed model aims to determine the time-dependent surface profile by solving both momentum and energy equations introduced in the previous paragraphs. It takes into account: (i). a solid-to-liquid phase transition with an energy transfer to the lattice equal to the latent heat (Eqs.1, 6), (ii). the Marangoni effect which describes liquid flow due to temperature gradients [34], (iii). the contribution of recoil pressure and pressure due to surface tension, and (iv) a resolidification process that is quantifiable by computing the liquid-solid interface velocity. Due to a small vertical anticipated surface modification with respect to the size of the laser beam, a finite difference method to solve the Eqs.(1-15) will suffice to produce accurate results. The coordinate system used in the analysis is defined as follows: the *z*-axis is normal to the material surface, the *x*-axis is on the surface with a direction based on that the electric field of the incident beam must reside on the *xz*-plane, and the *y*-axis, again on the material surface. Due to the axial symmetry (for single pulse irradiation), cylindrical coordinates (*r* and *z*) can be employed to obtain the carrier and lattice temperatures and the surface modification details. As a result, we can perform a simulation on a rectangular subregion of thickness $W=5\mu m$ and length $L=30\mu m$ is selected. The simulation runs for timepoints in the range from $t=0$ to 15ns in time intervals that vary from 10fs to 50fs to integrate nonequilibrium time history while a substantially larger time step (i.e. 5ps) is followed for post equilibrium evolution. Carriers and lattice temperatures are set to $T=T_0=300^0K$ (room temperature) at $t=0$ while the initial concentration of the carrier is set to $N=1\mu m^{-3}$ [44]. **Simulation is conducted for fluence $E_d=0.7J/cm^2$, pulse duration $\tau_p=430fs$, and spot radius $R_0=15\mu m$.**

The hydrodynamic equations will be solved in the aforementioned subregion that contains either solid or molten material. To include the 'hydrodynamic' effect of the solid domain, material in the solid phase is modelled as an extremely viscous liquid ($\mu_{solid}=10^5 \mu_{liquid}$), which will result into velocity fields that are infinitesimally small. An apparent viscosity is then defined with a smooth switch function similar to Eq.5 to emulate the step of viscosity at the melting temperature. A similar step function is also introduced to allow a smooth transition for the rest of temperature dependent quantities (i.e. heat conductivity, heat capacity, density, etc) between the solid and liquid phases. For time-dependent flows, a common technique to solve the Navier-Stokes equations is the projection method and the velocity and pressure fields are calculated on a staggered grid (Fig. 1a in [30]) using fully implicit formulations [61, 62]. The inset in Fig.(1a) illustrates the different locations at which the velocity ($v,u$) and pressure $P$ fields are calculated at a timepoint $t=t^l$. More specifically, the horizontal and vertical velocities are defined in the centres of the horizontal and vertical cells faces, respectively where the pressure and temperature fields are defined in the cell centres. Similarly, all temperature dependent quantities (i.e. viscosity, heat capacity, density, etc) are defined in the cell centres. An explicit solution method is employed to solve Eq.10 and compute temperature values at subsequent time points. The size of the horizontal side of the computation grid is 0.1nm while the vertical side is taken to be equal to 0.01nm.

During the ultrashort period of laser heating, heat loss from the upper surface of target is assumed to be negligible. As a result, the heat flux boundary conditions for carriers and lattice are zero throughout the simulation while a zero flux at $r=0$ must be also imposed. Furthermore, it is assumed that only the top surface is subjected to the Gaussian-shape laser beam of an irradiation spot-radius $R_0$=15μm. Regarding the momentum conditions at the boundaries, we impose the following constraints:

1. $u_r=0$, on the symmetry axis $r=0$,
2. $\vec{u} = 0$, on the solid-liquid interface (non-slipping conditions),
3. $\mu \dfrac{\partial u_r}{\partial z} = \dfrac{\partial \sigma}{\partial T_s} \dfrac{\partial T_s}{\partial r}$, on the upper flat free surface ($T_s$ is the surface temperature) [63], and

   $\mu \dfrac{\partial u_\tau}{\partial n} = \dfrac{\partial \sigma}{\partial T_s} \dfrac{\partial T_s}{\partial \tau}$, on the rest curved free surface [54], where $\tau$ and $n$ are the surface tangent direction and normal component, respectively.

The aforementioned boundary conditions are valid when the first incident pulse irradiates the initial planar surface. The first boundary condition excludes transverse component of the velocity on the symmetry axis ($r=0$). The last expression describes the shear stress which is due to the surface tension gradient and it is exerted on the free surface and it is important to incorporate the effect of Marangoni flow. The temperature-dependent parameters that are used in the numerical solution of the governing equations are listed in Table I. In order to increase accuracy, the model geometry is divided into two regions: a subdomain in which solid phase dominates and velocity fields are minimal and another subdomain where both momentum and energy equations are solved with an increased tolerance level.

For irradiation with subsequent pulses, we note that the incident beam is not always perpendicular to the modified profile, therefore the surface geometry influences the spatial distribution of the deposited laser energy. Hence, the laser irradiation reflected from the profile slopes can lead to light entrapment among the formed structures where the laser fluence is modified. Typical Fresnel equations are used to describe the reflection and transmission of the incident light. Due to multiple reflection, absorption of the laser beam is modified [64] and thereby a ray tracing method is employed to compute the absorbed power density while a similar methodology is ensued to estimate the proportion of the refraction by applying Snell's law. With respect to the numerical scheme used to simulate dynamics of velocity and pressure fields and all thermophysical quantities in Eqs.(1-12), a similar procedure is ensued in the event of subsequent pulses, however in this case the interaction with the modified surface profile induced by the first pulse due to the hydrodynamic motion of the molten material and its subsequent resolidification, should be taken into account. While second order finite difference schemes appear to be accurate for $NP$=1 where the surface profile has not been modified substantially, finer meshes and higher order methodologies are performed for more complex and profiles [65, 66]. The calculation of the pressure associated to the surface tension requires the computation of the temporal evolution of the principal radii of surface curvature $R_1$ and $R_2$ that correspond to the convex and concave contribution, respectively [67]. Hence the total curvature is computed from the expression $K=(1/R_1 +1/R_2)$. A positive radius of the melt surface curvature corresponds to the scenario where the centre of the curvature is on the side of the melt relative to the melt surface.

Regarding the material removal simulation, in each time step, lattice and carrier temperatures are computed and if lattice temperature reaches ~$0.90T_{cr}$, mass removal through evaporation is assumed. In that case, the associated nodes on the mesh are eliminated and new boundary conditions of the aforementioned form on the new surface are enforced. In order to preserve the smoothness of the surface that has been removed and allow an accurate and non-fluctuating value of the computed curvature and surface tension pressure, a fitting methodology is pursued [30].

3. **Surface pattern with ripples.**

To investigate incubation effects and the evolution of the ripple periodicity, the surface profile modification is probed after irradiating the material with more pulses. A systematic analysis of the fundamental mechanism reveals that inhomogeneous (i.e. periodic) deposition of the laser energy leads to recoil pressure, surface tension variance and temperature gradients leads to a periodic distribution of the lattice temperature which results into capillary-driven molten material flow (Fig.1a) [29] (for $NP$=10). Upon solidification, the resulting ripple periodicity varies according to the number of times the material is being irradiated (Fig.1b) [30].

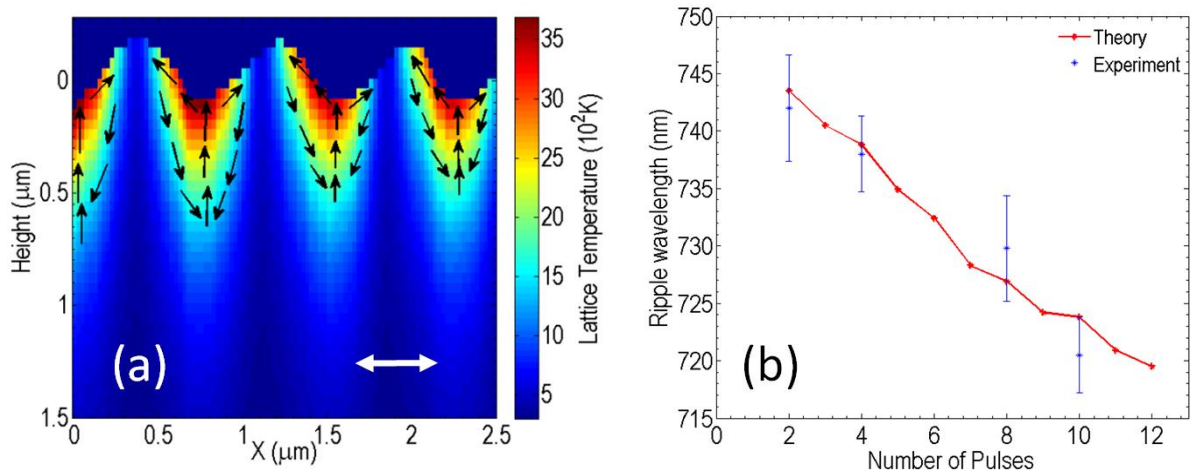

FIG.1. (a) Spatial distribution of lattice temperature at $t$=1ns for $NP$=10 (Arrows indicate flow movement), (b) Ripple periodicity vs. number of pulses. ($E_d$=0.7J/cm$^2$, $\tau_p$=430fs, $R_0$=15μm). (Double-ended arrow indicates the laser beam polarisation).

Fig.2a illustrates the three dimensional spatial dependence of the surface patterning for $NP$=10 in one quadrant where there is a pronounced rippled surface. A cross section (*dashed* line in Fig.2a) across a ripple indicates a spatial decrease of the ripple's amplitude followed by the occurrence of a small protrusion at the edge of the affected region (Fig.2b). The average ripple period after laser irradiation with *NP*=10 pulses is estimated to be equal to ~724nm. For a detailed account of the dependence of the ripples' periodicity on the number of pulses see [30].

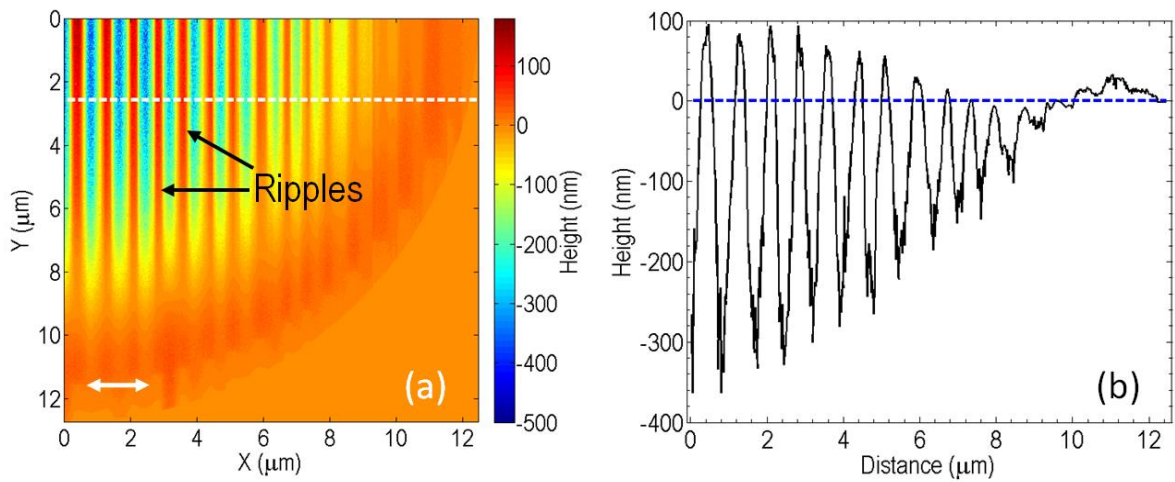

FIG.2. Theoretical results for: (a) Upper view of ripple pattern in a quadrant for *NP*=10 (double-ended arrow indicates beam polarisation), (b) Height of ripples (side view) for *NP*=10 at *Y*=2.5μm (along the *dashed* line in (a)). ($E_d$=0.7J/cm$^2$, $\tau_p$=430fs, $R_0$=15μm).

4. **Efficacy factor.**

The inhomogeneous energy deposition into the irradiated material is computed through the calculation of the product $\eta(\boldsymbol{k},\boldsymbol{k}_i) \times |b(\boldsymbol{k})|$ as described in the Sipe-Drude model [38, 39]. In the above expression, $\eta$ describes the efficacy with which the surface roughness at the wave vector $\boldsymbol{k}$ (i.e. $|\boldsymbol{k}|=2\pi/\lambda$) induces inhomogeneous radiation absorption, $\boldsymbol{k}_i$ is the component of the wave vector of the incident beam on the material's surface plane and $b$ represents a measure of the amplitude of the surface roughness at $\boldsymbol{k}$. The efficacy factor depends on the dielectric constant $\varepsilon$ of the material [39] and therefore any variation of the carrier temperature and density due to inhomogeneous radiation absorption is expected to influence the dielectric constant and SPW periodicity.

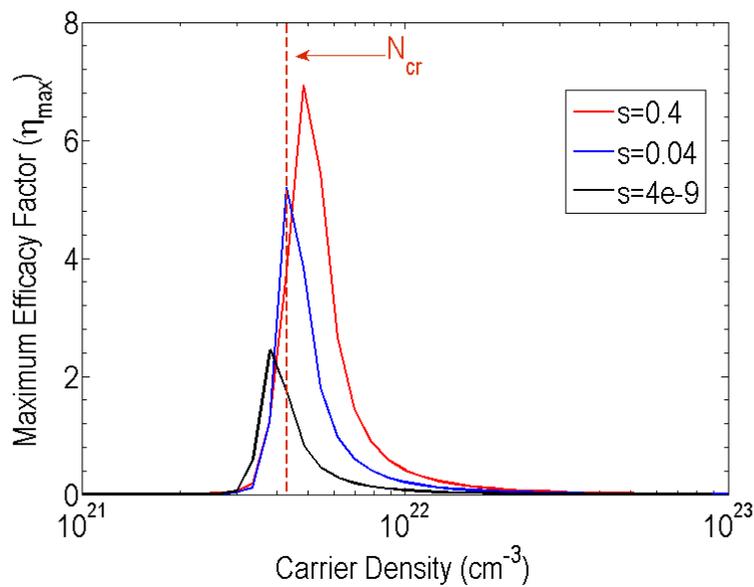

FIG.3. Efficacy factor for various values of *b* ($E_d$=0.7J/cm$^2$, $\tau_p$=430fs, $R_0$=15μm).

Fig.3 illustrates the computed value of the efficacy factor as a function of the carrier density for various values of the shape factor, *s* [38, 39] which characterizes the shape of the structures that make up the corrugated surface. It appears that for smaller number of pulses where ripples are not very sharp (i.e. smaller amplitudes of the ripples and larger *s*), the efficacy factor attains larger values. (For comparison reasons, we have used *s*=0.4 that corresponds to spherical-shaped islands [25, 39]). By contrast, when the shape of the ripples starts to become very sharp which occurs for *NP*~20, *s* is decreasing (*s*→0 for spike or 'banana shape' [38, 39]) which suggests a decreasing energy deposition and a smaller ripple amplitude. As a result, the efficacy with which the rough surface absorbs energy is reduced expected ripples' amplitude diminishes which will also result to gradually disappearance of the ripples. This is also justified by the fact that: (i) the maximum peak of the efficacy factor reduces to values that correspond to carrier densities lower than $N_{cr}$ (=4.29x10$^{21}$cm$^{-3}$) which is the threshold for SPW excitation, and (ii) for smaller values of the shape factor, the efficacy factor (energy absorption) is increasingly small for larger carrier density values.

5. **Fluid Mechanics and formation of hydrothermal waves.**

In this section we will provide the fundamentals of the fluid mechanics mechanisms that lead to formation of hydrothermal waves based on previous works(see [36, 40, 41, 43, 68] and references therein) considering also a revision due to the inclined profile resulted from the ripple geometry [69].

We consider an incompressible fluid with small variations *ΔT* about a constant value $T_0$ leading to variations in the local fluid density (Boussinesq approximation that assume buoyancy forces)

$$T = T_0 + \Delta T$$
$$\rho = \rho_0 - \alpha \rho_0 \Delta T \quad (16)$$

where *α* is the coefficient of thermal expansion of the material and *ρ* is taken constant (*ρ=ρ₀*) everywhere except the body force in the following equation

$$\rho_0\left(\frac{\partial \vec{u}}{\partial t}+\vec{u}\cdot\vec{\nabla}\vec{u}\right)=\vec{\nabla}\cdot\left(-P\mathbf{1}+\mu\left(\vec{\nabla}\vec{u}\right)+\mu\left(\vec{\nabla}\vec{u}\right)^T\right)+\rho_0(1-\alpha\Delta T)\vec{g} \qquad (17)$$

where $\vec{g}$ is the acceleration of gravity. In a simplified scenario in which temperature gradient is assumed the solution of Eqs.(7,8,10,17) is performed through a linear stability analysis. We start from an initial flow (i.e. a base state (BS) with $\vec{u}_{BS}=0$) that represents a stationary state of the system. Then, we subject the base state to an infinitesimal perturbation for the various physical variables (i.e. velocity, temperature and pressure)

$$\begin{aligned}\vec{u}&=\vec{u}_{BS}+\delta\vec{U}_p\\ T&=T_{BS}+\delta T_p\\ P&=P_{BS}+\delta P_p\end{aligned} \qquad (18)$$

where $\partial_z P_{BS}=-g\rho_0\left[1+\alpha\partial_z T_{BS}z\right]$ and we obtain the equations of motion by neglecting terms $O(\delta^2)$. $\delta$ stands for an infinitesimal constant coefficient that is used to ensure that the additional term is a perturbation. Calculations and reorganization of the terms in the equations yield (for a temperature gradient along the $z$-axis)

$$\begin{aligned}\partial_t\vec{\nabla}^2 U_p^{(z)}&=\alpha g\left(\partial_x^2+\partial_y^2\right)T_p+(\mu/\rho_0)\vec{\nabla}^4 U_p^{(z)}\\ \partial_t T_p-\left(\partial_z T_{BS}\right)U_p^{(z)}&=K\vec{\nabla}^2 T_p\end{aligned} \qquad (19)$$

where the initial problem has been reduced to a coupled system of partial differential equations that contain the perturbation for the vertical (i.e. $z$-axis) component of the velocity, $U_p^{(z)}$, and the temperature $T_p$.

The form of the Eqs.19 allows to introduce separable normal mode solutions $W(z)e^{-\omega t}\sum_n c_n e^{i\vec{k}_n\cdot\vec{r}}$ and $\theta(z)e^{-\omega t}\sum_n c_n e^{i\vec{k}_n\cdot\vec{r}}$ for the velocity (along $z$-direction) and the temperature component and solve the produced partial differential equations semi-analytically or numerically by applying appropriate boundary conditions depending on whether boundaries are no-slip or stress free [41, 43, 70]. The solution allows to determine the condition for which $\omega<0$ and subsequently Marangoni or Rayleigh number as a function of a wavenumber $|k|$ that produces the lowest Marangoni or Rayleigh number at which a disturbance with this periodicity becomes unstable (see [41, 43, 70]). The wavenumber that leads to the critical Marangoni or Rayleigh numbers constitutes the wavenumber that corresponds to the periodicity at the onset of the hydrothermal convection.

To describe the movement and convection of the fluid in the case that temperature gradient is not vertical, an appropriate modification is required to take into account that all thermo-physical quantities that appear in the heat transfer and Navier-Stokes equations have a spatio-temporal dependence. Fig.2b in the main manuscript shows that there is a gradient of the temperature with a small contribution along the *Y*-axis (i.e. $T_{BS} \equiv T_{BS}(x,z)$). Given the variable depth of the crater, we introduce a local Cartesian coordinate system to describe movement on an inclined layer which is characterized by a variable angle $\gamma \equiv \gamma(x,y,z)$. A position in the new coordinate system is characterized by ($X' \equiv x$, $Y'$, $Z'$) where the unit vectors are given by the expressions

$$\hat{Y}' = \cos(\gamma)\hat{y} + \sin(\gamma)\hat{z}$$
$$\hat{Z}' = -\sin(\gamma)\hat{y} + \cos(\gamma)\hat{z} \qquad (20)$$

where the original Cartesian coordinate system is characterised by the unit vectors $\hat{x}, \hat{y}, \hat{z}$. Assuming only steady motion the heat transfer equation yields

$$\frac{d^2 T_{BS}(X',Z')}{d(X')^2} + \frac{d^2 T_{BS}(X',Z')}{d(Z')^2} = 0 \qquad (21)$$

while the *X'*, *Z'* of Eq.19 give (considering that $\vec{g} = g\left(-\sin(\gamma)\hat{Y}' + \cos(\gamma)\hat{Z}'\right)$ and $\vec{u}_{BS} \equiv \vec{u}_{BS}(x,Z')$)

$$\frac{1}{\rho_0}\frac{\partial P_{BS}(X',Y',Z')}{\partial X'} = (\mu/\rho_0)\left[\frac{d^2 u^{(x)}_{BS}(X',Z')}{d(X')^2} + \frac{d^2 u^{(x)}_{BS}}{d(Z')^2}\right]$$

$$\frac{1}{\rho_0}\frac{\partial P_{BS}(X',Y',Z')}{\partial Y'} = g\alpha T_{BS}(X',Z')\sin(\gamma) + (\mu/\rho_0)\left[\frac{d^2 u^{(y)}_{BS}(X',Z')}{d(X')^2} + \frac{d^2 u^{(y)}_{BS}}{d(Z')^2}\right] \quad (22)$$

$$\frac{1}{\rho_0}\frac{\partial P_{BS}(X',Y',Z')}{\partial Z'} = g\alpha T_{BS}(X',Z')\cos(\gamma) + (\mu/\rho_0)\left[\frac{d^2 u^{(z)}_{BS}(X',Z')}{d(X')^2} + \frac{d^2 u^{(z)}_{BS}}{d(Z')^2}\right]$$

Solving numerically the above PDEs leads to the derivation of the spatial dependence of the base state for pressure, velocity and temperature. Then, we subject again as in the simpler case, the base state to an infinitesimal perturbation for the various physical variables (i.e. velocity, temperature and pressure)

$$\vec{u} = \vec{u}_{BS} + \delta\vec{U}_p$$
$$T = T_{BS} + \delta T_p \quad (23)$$
$$P = P_{BS} + \delta P_p$$

to obtain the equations of motion by neglecting terms $O(\delta^2)$. $\delta$ stands for an infinitesimal constant coefficient that is used to ensure that the additional term is a perturbation. As previously, the perturbation of velocity, temperature and pressure will be introduced in the form of periodic functions (normal modes) and the system of the PDEs is solved to ensure that $\omega<0$. These conditions allow to determine the critical value of the Marangoni number and the associated wavenumber that yields instability and **hydrothermal wave formation perpendicularly to the defined by the ripples' profile (i.e. parallel to the polarization vector).** Fig.4a illustrates a contour plot of the stream function (i.e. stream function $\psi$ is computed by solving the Poisson equation $\vec{\nabla}^2\psi = \vec{\nabla}\times\vec{U}$, where $\vec{U}$ is the velocity at every point) at $t=1$ns for $NP=20$ along $Y$-axis. Due to the opposite direction of rotation of the two rolls, the vorticity $\vec{\nabla}\times\vec{U}$ drops significantly at the meeting point of the two waves. Note that the *absolute* value of the stream function field for the convection roll on the right was considered (negative values is demonstrated by the opposite direction of flow (counterclockwise)). The velocity field vector is also illustrated in Fig.4b where the velocity vectors are shown on the stream function contour plot

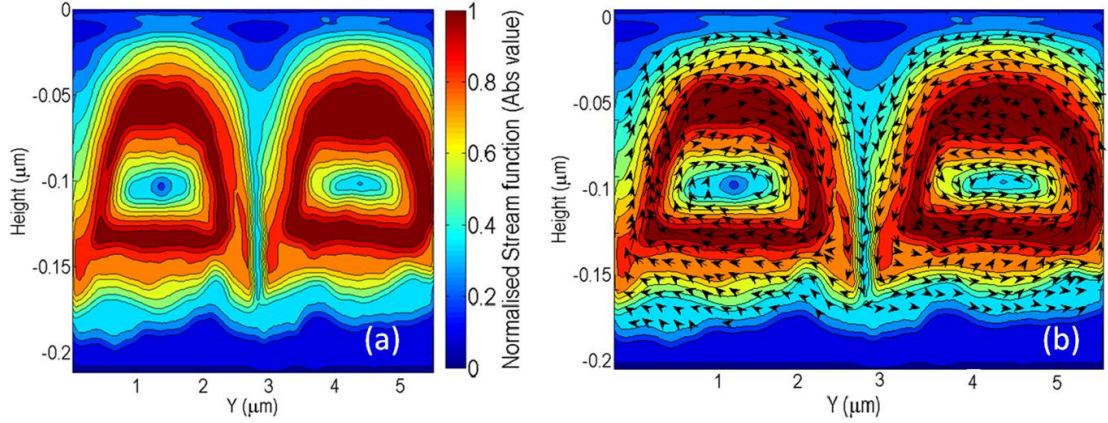

FIG.4. (a) Isolines plot for stream function and, (b) Velocity field for $NP$=20 at $t$=1ns ($E_d$=0.7J/cm$^2$, $\tau_p$=430fs, R$_0$=15μm).

## 6. Grooves.

The height profile was sketched after irradiation with $NP$=20 pulses to illustrate the morphological patterning induced by the flow movement of the material. More specifically, three dimensional simulations suggest that for $NP$=20, grooves and ripples are formed (Fig.5a). A closer look on the groove surface is examined by taking a cross profile along the $X$-axis that shows that the groove is corrugated (i.e. formation of pseudoripples) while conventional ripples are also produced outside the groove where laser beam energy deposition is lower (Fig.5b). It is evident that the amplitude of the periodically corrugated region is substantially smaller that the amplitude of the conventional ripples. Moreover, a cross line along the $Y$-axis illustrates the periodicity of the groove and the absence of similar corrugation (i.e. the small deviations should not be attributed to any significant effect but minor hydrodynamical instabilities and computational errors could explain them) (Fig.5c).

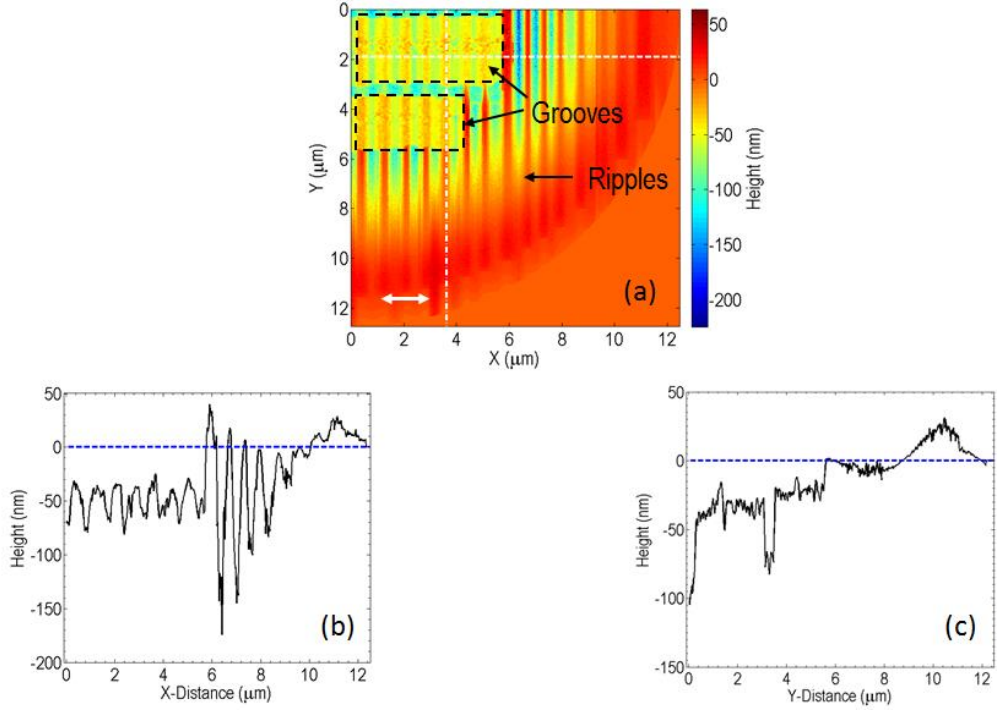

FIG.5. (a) Groove-ripple pattern for *NP*=20, (b) Height profile along *X*-axis at *Y*=1.92μm (*white dashed* horizontal line), (c) Height profile along *Y*-axis at *X*=3.63μm (*white dotted* horizontal line),. ($E_d$=0.7J/cm$^2$, $\tau_p$=430fs, $R_0$=15μm). (Double-ended arrow indicates the laser beam polarisation).

### 7. Hill/Spikes formation.

The height profile was also sketched after irradiation with *NP*=120 pulses to illustrate the morphological patterning induced by the flow movement of the material for a large number of pulses. More specifically, three dimensional simulations suggest that for *NP*=120 small protrusions (hills), grooves, pseudoripples and ripples are formed (Fig.6a). A closer look on the groove surface is examined by taking a cross profile along the *X*-axis that shows that the groove is corrugated (i.e. formation of structures that are different from the pseudoripples) while similar structures described in the previous section are also produced (Fig.6b). It is evident that the periodicity and the amplitude of the small hills is increased compared to the periodicity and the amplitude of the pseudoripples. Furthermore, a cross line along the *Y*-axis illustrates clearly the height of the small hills on top of the grooves that can be viewed as the precursors of spikes (Fig.6c).

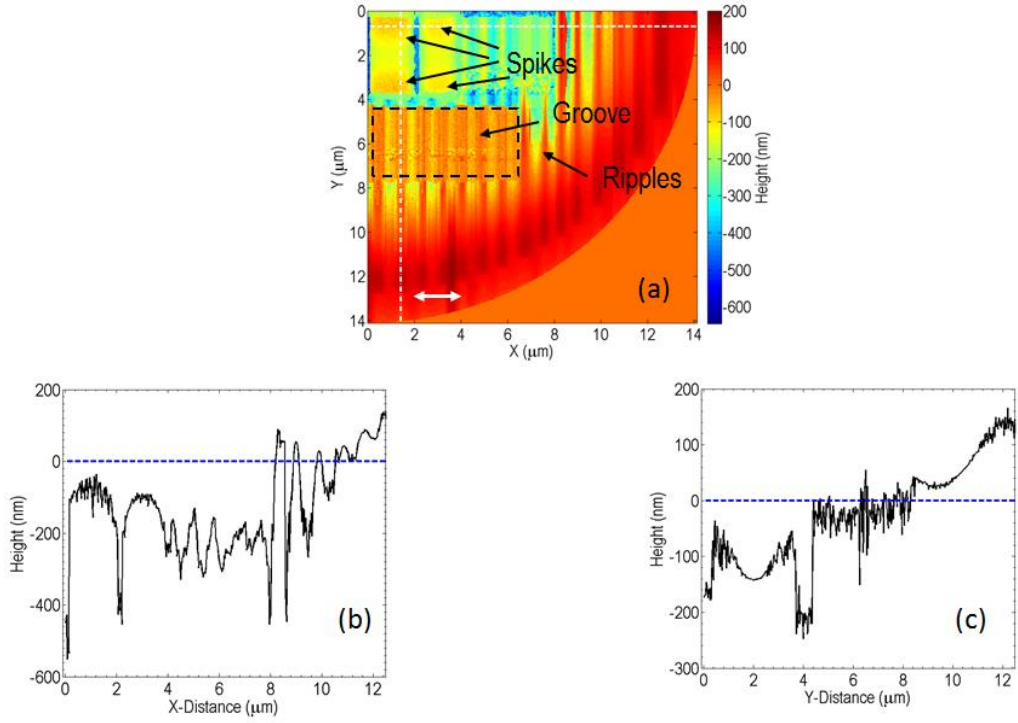

FIG.6. (a) Spike-groove-ripple pattern for *NP*=120, (b) Height profile along *X*-axis at *Y*=0.71μm (*white dashed* horizontal line), (c) Height profile along *Y*-axis at *X*=1.42μm (*white dotted* horizontal line),. ($E_d$=0.7J/cm$^2$, $\tau_p$=430fs, $R_0$=15μm). (Double-ended arrow indicates the laser beam polarisation).

## 8. **Ripples and Grooves in Titanium.**

To validate the theoretical results that indicate the presence of 'ripple-like' (or pseudoripples) on top of the grooves, SEM images collected during laser irradiation of Titanium with a moderate number of pulses (*NP*=50). Fig.7a provides a view of the coexistence of ripples in the periphery of the spot and grooves closer to the centre. A region (inside the *green* boundary) that contains grooves, illustrates clearly (Fig.7b) the presence of pseudoripples on the grooves while an intensity image profile analysis (Fig.7c) along a groove (*red* line) provide details with respect to their periodicity (i.e. equal to the ripples' periodicity).

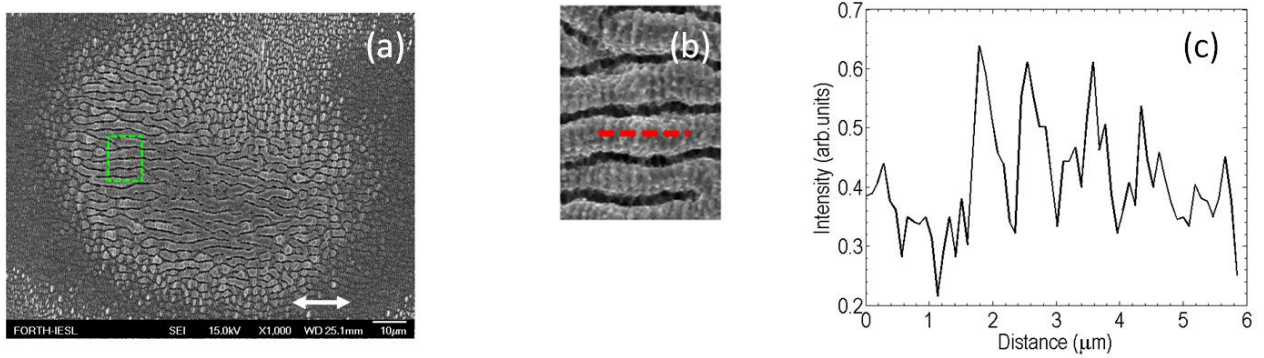

FIG.7. (a) SEM image after Titanium was irradiated with 50pulses (double-ended arrow indicates beam polarisation), (b) Magnified picture of region inside the *green* boundary in (a), (c) Intensity profile along the groove (r*ed* line in (b)).

9. **Experimental details.**

Experiments were performed with a femtosecond Ti:Sapphire laser system operating at a wavelength of 800 nm and repetition rate of 1 kHz. The pulse duration was set to 430 fs and measured by means of cross correlation techniques. A pockels cell controlled the repetition rate and the number of the pulses that irradiated the silicon surface. The beam was perpendicular to the silicon substrate located inside a vacuum chamber evacuated down to a residual pressure of $10^{-2}$ mBar. The laser fluence calculated from the beam waist ($1/e^2$) was 0.7 J/cm$^2$, chosen to assure a minimal mass removal for a single pulse irradiation. This is performed by means of atomic force (AFM- Nanonics Multiview-4000) and Field emission scanning electron microscopy (FESEM – JEOL JSM-7000F) images of the profiles of the various modification spots obtained.

The authors acknowledge support from the *'3DNeuroscaffolds'* research project.